\documentclass[prd,showpacs,showkeys,preprint]{revtex4}

\usepackage{amsmath}
\usepackage{amssymb}
\usepackage{yfonts}[1988/10/03]
\usepackage{graphicx}
\newcommand {\beq}{\begin{equation}}
\newcommand {\eeq}{\end{equation}}
\newcommand {\beqa}{\begin{eqnarray}}
\newcommand {\eeqa}{\end{eqnarray}}

\begin{document}
\title{ Damping of tensor mode in spatially closed cosmology } 
\author{Jafar Khodagholizadeh}
\email{j.gholizadeh@modares.ac.ir}
% \altaffiliation[Physics Department, School of science,Tarbiat Modares University, P.O.box 14155-4838,Tehran,Iran ]{}%Lines break automatically or can be forced with \\
\author{Amir H.Abbassi}%
 \email{ahabbasi@modares.ac.ir}
 \author{Ali A. Asgari}
 \email{aliakbar.asgari@modares.ac.ir}
\affiliation{%
Physics Department, School of sciences,Tarbiat Modares University, P.O.box 14155-4838,Tehran,Iran
}%

\begin{abstract}
 We derive an integro-differential equation for propagation of cosmological gravitation waves in spatially closed cosmology whereas the traceless transverse tensor part of the anisotropic stress tensor is free streaming neutrinos (including antineutrinos), which have been traveling essentially without collision since temperature dropped below about $ 10^{10} K$. We studied the short wavelengths and long wavelengths of gravitational waves (GWs) that enter the horizon in closed spacetime. The solution shows that the anisotropic stress reduces the squared amplitude by $ 76\%$ for wavelengths that enter the horizon during radiation-dominated phase and this reduction is less for the wavelength that enter the horizon at later times. At the end we compare the results to the flat case and then we investigated the dependence of the evolution of GWs on cosmological parameters.

\end{abstract}
%\preprint{***}
\maketitle

\section{Introduction}
The BICEP2 collaboration reports the signal of gravitational waves in the B-mode power spectrum in spatially flat spacetimes\cite{BICEP2}. These tensor perturbations (gravitational waves) arises from infaltionary in the early universe. As is well known, gravitational waves propagate freely in the expanding universe with this assumption that the Universe is perfect fluid \cite{Lifshitz} and also the cosmological generation of gravitational waves was considered by L.P.Grishchuk \cite{Grishchuk} and A.A.Starobinsky\cite{Starobinsky}. Even though in WMAP nine-year data report the results have been interpreted as an evidence in the support of flat universe, in no way do the data rule out the case of $ K=1 $\cite{ komatu}.   By choosing the maximally extended de-Sitter metric as unperturbed background, we found the evolution of tensor modes fluctuation  in curved spacetime \cite{gholizadeh} .   We compute only the tensor mode fluctuation between the time of inflation and present specially from a temperature ~$  \approx 10^{10}K $ when electron positron annihilation is substantially complete and neutrinos have decoupled from matter and radiation.
This paper is organized as follows. In Sec.II, we present the neutrino contribution to the anisotropic inertia tensor $ \Pi_{ij}^{T} $ so that for calculation of high accuracy it is necessary to use the Boltzmann equations of kinetic theory for neutrinos in phase space  and then get the integro-differential equation of tensor mode fluctuation in closed cosmology. In Sec.III, we discuss short wavelengths re-entering the horizon during the radiation dominated phase because modes are time-independent when leaving the horizon. In Sec.IV, we present the GW's equation in matter dominated era and discuss the effects of free streaming neutrinos on it. In last section we conclude with discussing and comparing the results of flat \cite{damping},\cite{Stefanek} and closed cosmology.
\section{DAMPING EFFECTS IN THE WAVE EQUATION}
In general the tensor mode gravitational perturbation  takes the form
\begin{eqnarray}
\delta g_{ij}=h_{ij}(\vec{x},t)=a^{2}(t) D_{ij}(\vec{x},t)
\end{eqnarray}
and the tensor fluctuation in curved spacetime satisfies\cite{gholizadeh}
\begin{eqnarray}\label{4}
\nabla^{2}D_{ij}-a^{2}\ddot{D}_{ij}-3a\dot{a}D_{ij}-2KD_{ij}=16\pi G\Pi_{ij}
\end{eqnarray}
for which the dot stands for derivative with respect to the ordinary time, $ K $ is curvature constant and $ \Pi_{ij} $ is anisotropic inertia tensor. In the perfect fluids and independent of the background space time the interaction of tensor modes with matter and radiation  is zero \cite{weinberg2,Rub, Fabbri,Abbott,Starobinskii}, but there is a traceless transverse term in the anisotropic tensor so $   \Pi_{ij} $ is only the anisotropic part of the stress tensor that is the spatial part of the perturbed energy-momentum tensor $ T_{ij}=\bar{p}g_{ij}+a^{2}\Pi_{ij} $ or
\begin{equation}
T^{i}_{~j}=\bar{p}\delta_{ij}+\Pi^{i}_{~j}-Kx^{i}x^{k}\Pi_{kj}
\end{equation}
where $ \bar{p} $ is unperturbed pressure. The components of the perturbed metric  in Cartesian coordinate are \cite{weinberg}
\begin{eqnarray}
g_{00}=-1~,~g_{i0}=0~,~g_{ij}=a^2(t)(\delta_{ij}+K\frac{x^ix^j}{1-Kx^2}+D_{ij}(\vec{x},t))
\end{eqnarray}
if we consider only tensor perturbation so that
\begin{eqnarray}
D_{ii}=0~~~,~~~ \nabla^{i}D_{ij}=0~~~,~~~\Pi_{ii}=0~~~,~~~\nabla^{i}\Pi_{ij}=0
\end{eqnarray}
The anisotropic inertia tensor is the sum of the contributions from photons and neutrinos but photons have a small contribution to the anisotropic inertia due to a short mean free time and neutrinos traveling without collisions when temperature dropped about  $ T=10^{10} K $ so  neutrino distribution function in phase space has a form
\begin{equation}
n_{\nu}(\vec{x},\hat{p},t)\equiv \sum_{r}\prod_{i=1}^{3}\delta^{(3)}(x^{i}-x_{r}^{i}(t))\prod_{i=1}^{3}\delta^{(3)}(p_{i}-p_{ri}(t))
\end{equation}
Where $  r $ is individual neutrinos and anti neutrino trajectories. In the absence of collisions terms, Boltzmann equation for neutrinos would be
\begin{eqnarray}
\dfrac{\partial n_{\nu}}{\partial t}+\dfrac{\partial n_{\nu}}{\partial x^{i}}\dot{x}^{i}+\dfrac{\partial n_{\nu}}{\partial p^{i}}p_{i}^{0}=0
\end{eqnarray}
where $ \dot{p}_{ri}=\dfrac{1}{2p_{r}^{0}}p_{r}^{j}p_{r}^{k}(\dfrac{\partial g_{jk}}{\partial x^{i}})_{x=x_{r}} $  and $ \dot{x}_{r}^{i}=\dfrac{p_{r}^{i}}{p_{r}^{0}} $  are the rate of the change of momentum  and the rate of change of the coordinate respectively, so the above relation will be
\begin{eqnarray}\label{1}
\dfrac{\partial n_{\nu}}{\partial t}+\dfrac{\partial n_{\nu}}{\partial x^{i}}\dfrac{p^{i}}{p^{0}}+\dfrac{\partial n_{\nu}}{\partial p^{i}}\dfrac{p^{j}p^{k}}{2p^{0}}\dfrac{\partial g_{jk}}{\partial x^{i}}=0
\end{eqnarray}
$ n_{\nu}(\vec{x},t) $ in the start of free streaming has the form of the ideal gas:
\begin{eqnarray}
\bar{n}_{\nu}(\vec{x},t)=\dfrac{N}{(2\pi)^{3}}/[exp(\dfrac{\sqrt{g^{ij}p_{i}p_{j}}}{k_{B}a(t)\bar{T}(t)})+1]
\end{eqnarray}
Where $ N $ is the number of types of neutrinos and separately antineutrinos
and $ k_{B} $ is the Boltzmann constant. With a small perturbation to the metric, the neutrino disturbation function gets a small difference from its equilibrium form as
\begin{eqnarray}
n_{\nu}(\vec{x},t)=n_{\nu}(a(t)\sqrt{g^{ij}p_{i}p_{j}})+\delta n_{\nu}(\vec{x},t)
\end{eqnarray}
 If the background spacetime is  non-flat$ (K\neq 0) $ so $ p^{i} $ , $ p $ and $ p^{0} $ are of independent variable $ p_{i} $ by $p^{i}=g^{ij}p_{j}=a^{-2}(p_{i}-Kx^{i}x^{j}p_{j})  $ and $p=\sqrt{\tilde{g}^{ij} p_{i} p_{j}}=\sqrt{(\delta^{ij}-Kx^{i}x^{j})p_{i}p_{j}}$ and $ p^{0}=\sqrt{g^{ij}p_{i}p_{j}} $.  To first order in metric and density perturbation Eq.(\ref{1}) will be
\begin{eqnarray}
\dfrac{\partial\delta n_{\nu}}{\partial t}+\dfrac{p^{i}}{a(t)p}\dfrac{\partial\delta n_{\nu}}{\partial x^{i}}+K \dfrac{\hat{p}_{i}}{a(t)}p x^{l}\hat{p}_{l}\dfrac{\partial\delta n_{\nu}}{\partial p_{i}}=\dfrac{p}{2}\hat{p}_{i}\hat{p}_{j}\bar{n}_{\nu}^{\prime}(p)\dfrac{\partial}{\partial t}(a^{-2}\delta g^{ij}) \nonumber\\
-K \dfrac{\bar{n}_{\nu}^{\prime}}{a^{3}}p \hat{p}_{k}x^{m}\delta g^{km}-K \dfrac{\bar{n}_{\nu}^{\prime}}{a^{3}}p \hat{p}_{i}(x^{l}\hat{p}_{l})\hat{p}_{k}\delta g^{ki}+K^{2} \dfrac{\bar{n}_{\nu}^{\prime}}{a^{3}}p x^{i}(x^{l}\hat{p}_{l})^{2}\hat{p}_{k}\delta g^{ki}\nonumber\\ -K^{2} \dfrac{\bar{n}_{\nu}^{\prime}}{2a^{3}}p \hat{p}_{i}x^{j}x^{k}(x^{l}\hat{p}_{l})^{2}\dfrac{\partial}{\partial x^{i}}\delta g^{jk}+K^{3} \dfrac{\bar{n}_{\nu}^{\prime}}{2a^{3}} x^{i}x^{j}x^{k}(x^{l}\hat{p}_{l})^{3}\dfrac{\partial}{\partial x^{i}}\delta g^{jk}
\end{eqnarray}
with $ \delta g_{ij}=a^{2}D_{ij}(\vec{x},t) $, the relativistic Boltzmann equation for the perturbation $ \delta n_{\nu}(\vec{x},\vec{p},t) $ in curved spacetime will be
\begin{eqnarray}\label{2}
\dfrac{\partial\delta n_{\nu}}{\partial t}+\dfrac{p^{i}}{a(t)p}\dfrac{\partial\delta n_{\nu}}{\partial x^{i}}+K \dfrac{\hat{p}_{i}}{a(t)}p x^{l}\hat{p}_{l}\dfrac{\partial\delta n_{\nu}}{\partial p_{i}}=\dfrac{p}{2}\hat{p}_{i}\hat{p}_{j}\bar{n}_{\nu}^{\prime}(p)\dfrac{\partial}{\partial t}D^{ij}(\vec{x},t) ~~~~~~~~~~~~~~~~\nonumber\\
-K \dfrac{\bar{n}_{\nu}^{\prime}(p)}{a}p \hat{p}_{k}x^{m}D^{km}(\vec{x},t)-K \dfrac{\bar{n}_{\nu}^{\prime}(p)}{a}p (x^{l}\hat{p}_{l})\hat{p}_{i} \hat{p}_{k}D^{ki}(\vec{x},t)+K^{2} \dfrac{\bar{n}_{\nu}^{\prime}(p)}{a}p x^{i}(x^{l}\hat{p}_{l})^{2}\hat{p}_{k} D^{ki}(\vec{x},t)\nonumber\\ -K^{2} \dfrac{\bar{n}_{\nu}^{\prime}(p)}{2a(t)}p \hat{p}_{i}x^{j}x^{k}(x^{l}\hat{p}_{l})^{2}\dfrac{\partial}{\partial x^{i}}D^{jk}(\vec{x},t)+K^{3} \dfrac{\bar{n}_{\nu}^{\prime}(p)}{2a} x^{i}x^{j}x^{k}(x^{l}\hat{p}_{l})^{3}\dfrac{\partial}{\partial x^{i}}D^{jk}(\vec{x},t)\nonumber\\
\end{eqnarray}
We use a dimensionless intensity perturbation $ J $, defined by
\begin{eqnarray}
a^{4}(t) \bar{\rho}_{\nu}(t) J(\vec{x},\vec{p},t)\equiv N_{\nu}\int_{0}^{\infty}\delta n_{\nu}(\vec{x},\vec{p},t) 4\pi p^{3}dp
\end{eqnarray}
where $ N_{\nu} $ is the number of species of neutrinos and antineutrinos and $ \bar{\rho}_{\nu} \equiv N_{\nu} a^{-4}\int 4\pi \rho^{3}\bar{n}_{\nu}(p)dp$ and also  with $ p_{i}\frac{\partial}{ \partial p_{i}}=p\frac{\partial}{\partial p} $ the Boltzmann equation (\ref{2}) becomes
\begin{eqnarray}\label{3}
\dfrac{\partial}{\partial t}J(\vec{x},\hat{p},t)+\dfrac{\hat{p}_{i}}{a(t)}\dfrac{\partial}{\partial x^{i}}J(\vec{x},\hat{p},t)-4 K\dfrac{x^{l}\hat{p}_{l}}{a(t)} J(\vec{x},\hat{p},t)=-2 \hat{p}_{i}\hat{p}_{j}\dot{D}^{ij}(\vec{x},t)~~~\nonumber\\+K\dfrac{4}{a(t)}\hat{p}_{k}x^{m}D^{km}(\vec{x},t)+K\dfrac{4}{a(t)}  \hat{p}_{m} \hat{p}_{k} (x^{l}\hat{p}_{l})D^{km}(\vec{x},t)-4\dfrac{K^{2}}{a(t)}(x^{l}\hat{p}_{l})^{2}\bar{\rho}_{\nu}x^{i}\hat{p}_{k}D^{ki}(\vec{x},t)\nonumber\\+K^{2}\dfrac{2}{a(t)}\hat{p}_{i} x^{j}x^{k}(x^{l}\hat{p}_{l})^{2}\dfrac{\partial}{\partial x^{i}}D^{jk}(\vec{x},t)- K^{3}\dfrac{2}{a(t)}x^{i} x^{j}x^{k}(x^{l}\hat{p}_{l})^{3}\dfrac{\partial}{\partial x^{i}}D^{jk}(\vec{x},t)~~~~~~~~~~~~~~\nonumber\\
\end{eqnarray}
the above equation can be have a solution in the following form
\begin{eqnarray}
J(\vec{x},\hat{p},t)=\sum_{\lambda=\pm 2}\sum_{q}\dfrac{1-(\vec{q}.\vec{x})^{2}}{1-q^{2}}e^{iq\arccos(\vec{q}.\vec{x})}e^{ij}(\hat{q},\lambda)\beta(\hat{q},\lambda)\hat{p}_{i}\hat{p}_{j}\Delta_{\nu}^{T}(q,\hat{p}.\hat{q},t)
\end{eqnarray}
 and also we define $ D_{ij}(\vec{x},t) $ as
\begin{eqnarray}
D_{ij}(\vec{x},t) =\int d^{2}\hat{q}\sum_{\lambda=\pm 2}\sum_{q}\dfrac{1-(\vec{q}.\vec{x})^{2}}{1-q^{2}}e^{iq\arccos(\vec{q}.\vec{x})}e^{ij}(\hat{q},\lambda)\beta(\hat{q},\lambda)D_{q}(t)
\end{eqnarray}
where $ \beta(\vec{q},t) $ is a stochastic parameter for the single non-decaying mode with discrete wave number $ q $ and the helicity $ \lambda $ and $ e_{ij}(\hat{q},t) $ is the corresponding polarization tensor. With $ \hat{p}_{i}\hat{q}_{i}=\mu $ and working in $ x\ll  1$ , Eq.(\ref{3}) becomes an equation for $ \Delta_{\nu}^{(T)} $:
\begin{eqnarray}
\dfrac{\partial}{\partial t}\Delta_{\nu}^{T}(q,\mu,t)+\dfrac{iq\mu}{a(t)}\Delta_{\nu}^{T}(q,\mu,t)-4 K\dfrac{q}{a(t)}\dfrac{\partial}{\partial \mu}\Delta_{\nu}^{T}(q,\mu,t)=-2 \dot{D}_{q}(t)~~~~~~~~~~~~
\end{eqnarray}
and we can find a solution of above equation with Green function method as
\begin{equation}
\Delta_{\nu}^{T}(q,\mu,\tau)=\int d\tau d\mu^{\prime}G(\mu,\mu^{\prime},\tau,\tau^{\prime}) (-2\dot{D}_{q}(\tau^{\prime}))
\end{equation}
 the Green function is obtained as
\begin{eqnarray}
G(\mu,\mu^{\prime},\tau,\tau^{\prime})=\dfrac{i}{2 \pi }\Theta(\tau^{\prime}q-\tau q) e^{-i\tau q(\mu+2 \tau q)}e^{i\tau^{\prime} q(\mu^{\prime}+ 2\tau^{\prime}q)}
\end{eqnarray}
in this case $\Delta_{\nu}(q,\mu,\tau)  $ is
\begin{eqnarray}
\Delta_{\nu}(q,\mu,\tau)= \dfrac{i}{2 \pi }e^{-i\tau q(\mu+ 2 \tau q)}\int _{-1}^{+1}d\mu^{\prime} \int_{0}^{\pi} d\tau^{\prime}[-2\dot{D}_{q}(\tau^{\prime})] \Theta(\tau^{\prime}q-2\tau q) e^{i\tau^{\prime} q(\mu^{\prime}+ 2\tau^{\prime}q)}\nonumber\\
\end{eqnarray}
in the tensor mode, the only nonvanishing component is $ \delta T^{i}_{~\nu j} $:
\begin{eqnarray}
\delta T^{i}_{~\nu j}(\vec{x},t)&=&a^{-4}(t) \int d^{3}p ~\delta n_{\nu}(\vec{x},\vec{p},t)p \hat{p}_{i}\hat{p}_{j}\nonumber\\ &=&\bar{\rho}_{\nu}(t)\Sigma_{\lambda}  \int d^{3}q \beta (\vec{q},\lambda) e^{i\vec{q}.\vec{x}}e_{ij}(\hat{q},\lambda)\times \frac{1}{4}\int\frac{d^{2}\hat{p}}{4\pi}\Delta_{\nu}^{T}(q,\hat{p}.\hat{q},t)(1-(\hat{p}.\hat{q})^{2})^{2}
\end{eqnarray}
This is the neutrino contribution of the anisotropic inertia tensor $ \Pi^{T}_{ij} $:
\begin{eqnarray}
\Pi^{T}_{ij}&=&\frac{\bar{\rho}_{\nu}(t)}{4}\int\frac{d^{2}\hat{p}}{4\pi}\Delta_{\nu}^{T}(q,\hat{p}.\hat{q},t)(1-(\hat{p}.\hat{q})^{2})^{2}\nonumber\\&=&-\dfrac{\bar{\rho}_{\nu}(t)}{4}(\dfrac{1}{2\pi })e^{-2i\tau^{2} q^{2}} \dfrac{1}{\tau^{5} q^{5}}[(-16\tau^{2} q^{2}+48)\sin(\tau q)-48\tau q \cos(\tau q)] \nonumber\\&\times & \int _{-1}^{+1}d\mu^{\prime} \int_{0}^{\tau q} d\tau^{\prime}[-2\dot{D}_{q}(\tau^{\prime})] e^{i\tau^{\prime} q(\mu^{\prime}+ 2\tau^{\prime}q)}
\end{eqnarray}
in the right hand side, the second and third terms compared with the first term are too small so by ignoring them, the integro-differential equation of gravitational waves in the presence of inertia tensor of neutrinos becomes
\begin{eqnarray}\label{7}
\ddot{D}_{n}(t)+3\dfrac{\dot{a}}{a}D_{n}(t)+\dfrac{q^{2}}{a^{2}(t)}D_{n}(t)=-64\pi G\bar{\rho}(\tau)e^{-2i\tau^{2} q^{2}}\dfrac{16}{\pi q} \dfrac{\sin \tau q}{\tau^{3} q^{3}} \int_{0}^{\tau q} d\tau^{\prime}[\dot{D}_{q}(\tau^{\prime})]\dfrac{\sin \tau^{'}q}{\tau^{'}q} e^{2i\tau^{\prime 2} q^{2}}
\end{eqnarray}
the right hand side of the equation (\ref{7}) is more complicated than the case of flat spacetime. In the next section we will calculate the decay of gravitational waves in the radiation dominated era.
\section{SHORT WAVELENGTHS}
We found a complete solution of differential equation for tensor perturbation in the closed universe\citep{gholizadeh}. We begin by neglecting the anisotropic inertia tensor $ \Pi_{ij}^{T} $ and then we consider $ x\ll 1 $ so with this approximation, the field equation (\ref{4}) governing the Fourier components of the tensor components of $ D_{ij} $ becomes
\begin{equation}\label{6}
\ddot{D}_{q}(t)+3\frac{\dot{a}}{a}\dot{D}_{q}(t)+\frac{q^{2}}{a^{2}}D_{q}(t)=0
\end{equation}
where $ q^{2}=n^{2}-2 $ and $ n $ is discrete. To treat the evolution of $D_{q}(t)  $, it is convenient to change the independent variable $ t  $ to $ y \equiv \frac{a}{a_{EQ}}=\frac{\bar{\rho}_{M}}{\bar{\rho}_{R}} $, where $ a_{EQ} $ is the value of the Robertson-Walker scale factor at matter-radiation equality. From Friedmann equation we have
\begin{eqnarray}\label{10}
H_{EQ}\dfrac{dt}{\sqrt{2}}=\dfrac{ydy}{\sqrt{1+y+2\Omega_{K_{EQ}}y^{2}+2\Omega_{V_{EQ}}y^{4}}}
\end{eqnarray}
where $ H_{EQ} $ is the expansion rate at matter-radiation equality, $ \Omega_{V_{EQ}} $ and $ \Omega_{K_{EQ}} $ are vacuum and curvature energy respectively when radiation and matter are equal.

To investigate a short enough wavelength  to have re-entered the horizon during the radiation dominated era, we consider $ y=\frac{\bar{\rho}_{M}}{\bar{\rho}_{R}} \ll 1$ the equation(\ref{6}) with the change of variable from $ t $ to $ y $ is
\begin{eqnarray}
\dfrac{d^{2}}{d y^{2}}D_{n}(y)+(\dfrac{2}{y})\dfrac{d}{dy}D_{n}(y)+ \kappa^{2}D_{n}(y)=0
\end{eqnarray}
where $ \kappa=\dfrac{\sqrt{2 } n}{a_{EQ}H_{EQ}} $ and $ n $ is discrete number. For $ y\longrightarrow 0 $ the solution is $ D_{n}^{0} =constant $ and general solution is
\begin{eqnarray}
D_{n}=D_{n}^{0} \dfrac{\sin ky}{ky}
\end{eqnarray}
the solution is similar to the flat case but $ \kappa$ depends on discrete number $ n $ in closed universe. The fraction of the total energy density in neutrinos is
\begin{eqnarray}
f_{\nu}(y)=\frac{\Omega_{\nu}(\frac{a_{0}}{a})^{4}}{\Omega_{M}(\frac{a_{0}}{a})^{3}+\Omega_{R}(\frac{a_{0}}{a})^{4}+\Omega_{\Lambda}}=\frac{f_{\nu}(0)}{1+y+(\frac{\Omega_{\Lambda}\Omega_{R}^{3}}{\Omega_{M}^{4}})y^{4}}
\end{eqnarray}
where $ f_{\nu}(0)=\frac{\Omega_{\nu}}{\Omega_{\nu}+\Omega_{\gamma}}=0.40523 $ and the third term in the denominator is too small and can be ignored. In the presence of the anisotropic inertia tensor with change $ u=q \tau=q \int_{0}^{t} \dfrac{dt}{a(t)}=\dfrac{2qt}{a(t)}$ instead of $ t $ and using the Friedmann equation $ \dfrac{8\pi G \bar{\rho}}{3}=H^{2}=\dfrac{1}{4 t^{2}} $, the gravitational wave equation(\ref{7}) in radiation dominated era becomes
\begin{eqnarray}\label{8}
\dfrac{d^{2}}{d u^{2}}D_{n}(u)+(\dfrac{2}{u})\dfrac{d}{du}D_{n}(u)+D_{n}(u)=\dfrac{-384 f_{v}}{\pi u^{2}} e^{-2iu^{2}}\dfrac{\sin u}{u^{3}}\int_{0}^{u}e^{2iu^{'2}}\dfrac{\sin u^{'}}{u^{'}}\dfrac{d D_{n}(u^{'})}{d u^{'}}du^{'}
\end{eqnarray}
when $ u\ll 1 $ the  homogeneous solution is constant value of $ D_{n}^{0} $ then $ \dfrac{d D_{n}(u^{'})}{d u^{'}}=0$ so the gravitational waves equation becomes
\begin{eqnarray}
\dfrac{d^{2}}{d u^{2}}D_{n}(u)+(\dfrac{2}{u})\dfrac{d}{du}D_{n}(u)+D_{n}(u)=0
\end{eqnarray}
the solution is identical to the flat spacetime but for $ u\gg 1 $ when the tensor modes are deep inside the horizon, the integro-differential  equation (\ref{8}) becomes
\begin{eqnarray}\label{101}
\dfrac{d^{2}}{d u^{2}}D_{n}(u)+(\dfrac{2}{u})\dfrac{d}{du}D_{n}(u)+D_{n}(u)=\dfrac{35.75 f_{v}(0)}{\pi}\dfrac{\sin u}{u^{5}}
\end{eqnarray}
and its general solution is
\begin{eqnarray}
D_{n}(u)=(D_{n}^{0} -3Ci(2u)+\dfrac{0.76}{u^{2}})\dfrac{\sin u}{u}
\end{eqnarray}
where $ D_{n}^{0} $ is constant and $ Ci(2u) $ is the Cosine integral as $ Ci(2u) =\gamma +\ln(2u) +\int_{0}^{2u}\dfrac{\cos t -1}{t} dt $. Deep inside of horizon when $ u\gg 1  $, the right hand side of the eq. (\ref{101}) becomes negligible and the solution approaches a homogeneous solution as
\begin{eqnarray}
D_{n}(u)\longrightarrow \frac{\sin u}{ u}
\end{eqnarray}
 for large $ u ~(u\gg 1)$,  $ Ci(2u) $ and the third term of the general solution tend to zero  so $ D_{n}^{0}=1 $. Also a numerical solution of  eq.(\ref{101}) shows that $ D_{n}(u) $ follows the $ f_{v}(0) $ solution pretty accurately until $ u \approx 1 $ when the perturbation enters horizon [as compared with the solution $ \dfrac{\sin u}{u} $ for $ f_{v}(0) $] and thereafter the solution rapidly approaches to the $ 0.4910 \dfrac{\sin u}{u} $. So then the effect of neutrino damping reduces the tensor amplitude by the factor of $ 0.4910 $ in  the closed cosmology while in the flat case the factor was $ 0.8026 $ \cite{Dicus}. Hence the tensor contribution to the temperature multipole coefficient $ C_{l} $ and the whole of the $ ^{''}B-B^{''} $  polarization multipole coefficient  $ C_{lB}$ will be $ 76 \% $ less than they would be without damping due to free -streaming neutrinos. Therefore in the radiation dominated era and in closed cosmology, the amplitude of the gravitational waves in presence of neutrinos will be less than the flat case or neutrinos have a greater effect on the damping of gravitational waves in closed cosmology as shown in Fig.1.
 \begin{figure}
\includegraphics[scale=0.6]{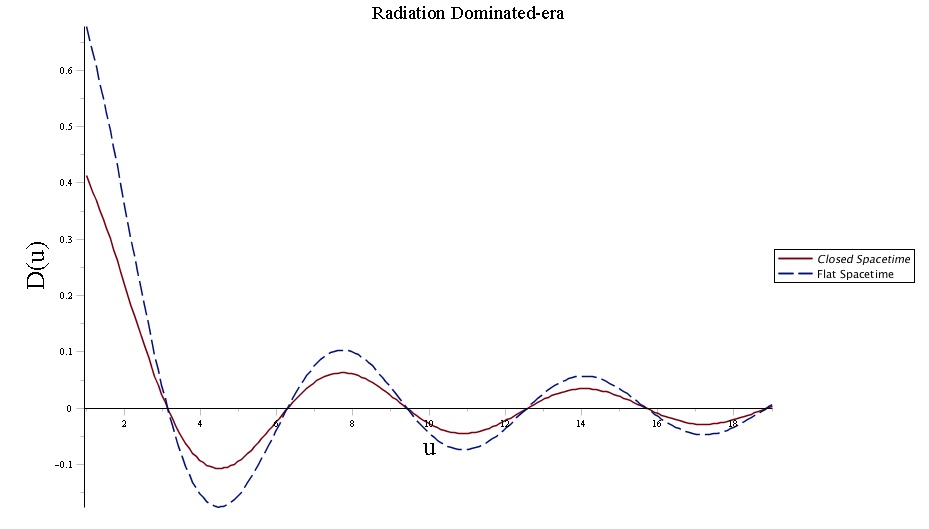}
\caption{In radiation dominated era when the perturbation enters the horizon,  the free-streaming neutrinos in closed cosmology (solid line) have a greater effect than the flat case(dashed line) on the damping of gravitational waves. }
\label{fig:1}
\end{figure}
\section{GENERAL WAVELENGTHS}
To investigate the tensor perturbation may enter horizon after the matter energy density becomes important, we consider $ y\gg 1 $ so from (\ref{10}) we have
\begin{eqnarray}
H_{EQ}\dfrac{dt}{\sqrt{2}}=\dfrac{ydy}{\sqrt{2\Omega_{V_{EQ}}y^{4}}}
\end{eqnarray}
where $ \Omega_{V_{EQ}} $ is vacuum energy at the equality of radiation and matter. Using $ \Omega_{V_{EQ}}=\frac{8\pi G \bar{\rho}}{3 H_{EQ}^{2}} $ and $ H=\sqrt{\frac{\Lambda}{3}} $ equation (\ref{6}) becomes
\begin{eqnarray}
\dfrac{d^{2}}{dy^{2}}D_{n}(y)+\dfrac{4}{y}\dfrac{d}{dy}D_{n}(y)+\dfrac{\kappa^{2}}{y^{4}}D_{n}(y)=0
\end{eqnarray}
where $ \kappa^{2}=\dfrac{n^{2}}{H_{EQ}^{2}a_{EQ}^{2}\Omega_{V_{EQ}}} $ and $ n $ is discrete number. This equation is different from gravitational waves equation in flat cosmology \cite{weinberg}. Whatever the value of $ \kappa $ , the general solution is
\begin{eqnarray}
D_{n}(u)=C_{n}^{1}[\dfrac{\kappa}{y}\cos(\dfrac{\kappa}{y})-\sin(\dfrac{\kappa}{y})]-C_{n}^{0}[\dfrac{\kappa}{y} \sin(\dfrac{\kappa}{y})+\cos(\dfrac{\kappa}{y})]
\end{eqnarray}
where $ C_{n}^{1} $ and $ C_{n}^{0} $ are constant. The solution is not oscillating in $k\ll 1$ and the neutrinos have no effect on the gravitational waves. Also we impose the condition that the solution should approach the constant of the value $ C_{n}^{0} $ for $\dfrac{k}{y} \ll 1$ so the solution will be
\begin{eqnarray}
D_{n}(u)=-C_{n}^{0}[\dfrac{\kappa}{y} \sin(\dfrac{\kappa}{y})+\cos(\dfrac{\kappa}{y})]
\end{eqnarray}
Similar to the flat spacetime, all wavelength to take the gravitational wave amplitude in matter-dominated era are to be given by multiplying the above solution with the factor $\alpha (k)$:
\begin{eqnarray}
D_{n}(u)=-C_{n}^{0}\alpha (k)[\dfrac{\kappa}{y} \sin(\dfrac{\kappa}{y})+\cos(\dfrac{\kappa}{y})]
\end{eqnarray}
where $\alpha (k)$ is $ 0.4910$ for $ k\gg 1$ and $\alpha (k)=1$ for $k\ll 1$ because the damping effect is equal to zero for $k\ll 1$. Therefore we can consider $\alpha (k)\cong \dfrac{1+0.4910 k}{1+k }$ and amplitude of gravitational waves will be reduced by the factor $\alpha (k)$.
\subsection{CONCLUSION}
At first we obtained the relativistic Boltzmann equation for perturbation $ \delta n_{\nu}(\vec{x}.\vec{p},t) $ in spatially closed cosmology, the perturbation on the metric background creates a disturbance in the number density of neutrinos and then we investigate the effects of anisotropic inertia tensor on the damping of gravitational waves in the radiation and matter dominated era. Amplitude of gravitational waves is a sensitive function of discrete wave number. During the radiation dominated era when the tensor perturbation re-entered the horizon, the effect of neutrinos on GWs in flat spacetime is weaker than the closed case and the evolution is independent of any cosmological parameters. In closed cosmology, in matter-dominated era the equation of gravitational waves is different from the flat case and the propagating of the gravitational waves will be different from the flat case and the effect is less for wavelengths that enter the horizon at later times. The evolution of GWs at later times after the radiation-matter equality depends on vacuum energy in the equality of matter and radiation.


\begin{thebibliography}{99}
  \bibitem{BICEP2}{ P.~A.~R.~Ade {\it et al.} [BICEP2 Collaboration], Phys.Rev.Lett.112,241101(2014)
[arXiv:1403.3985 [astro-ph.CO]].
}
\bibitem{Lifshitz} E.M. Lifshitz, Zh. Eksp.Teor.Phys. 16,587 (1946) ; L.P. Grishchuk,  Zh .Eksp. Teor. Fiz. 67 ,825 (1974)[Sov. Phys.JETP 40,409(1975)]; L.H. Ford and L.Parker, Phys. Rev.D 16.1601 (1977)
 \bibitem{Grishchuk}L.P. Grishchuk, JETP 40, 409(1974).
 \bibitem{Starobinsky}A.A.Starobinsky, JETP lett. 30, 682(1979).
\bibitem{komatu} E. Komatu and  et. al, APJS, 192: 18, (2011).
  \bibitem{gholizadeh}
 {   A.~H.~Abbassi, J.~Khodagholizadeh and A.~M.~Abbassi,
     %``Gravitational waves in a spatially closed de Sitter spacetime,''
   Eur.\ Phys.\ J.\ C {\bf 73}, 2592 (2013).
   [arXiv:1207.0876 [gr-qc]].
   %%CITATION = arXiv:1207.0876%%
   }

\bibitem{damping}
 S.~Weinberg,
 %``Damping of tensor modes in cosmology,''
Phys.\ Rev.\ D {\bf 69}, 023503 (2004).
[astro-ph/0306304].
%%CITATION = astro-ph/0306304%%
\bibitem{Stefanek}B.B.Stefanek, W.W.Repko,Phys.Rev.D ,88 ,083536(2013)
 \bibitem{weinberg2} S.Weinberg, Gravitation and Cosmology(Weily,New York,1972).
\bibitem{Rub}V. A. Rubakov, M. Sazhin, and A. Veryaskin, Phys. Lett. 115B, 189(1982).
\bibitem{Fabbri} R. Fabbri and M.D. Pollock, Phys. Lett. 125B, 445 (1983).
\bibitem{Abbott}L.F. Abbott and M. B. Wise, Nuclear Physics B244, 541 (1984).
\bibitem{Starobinskii}A. A.Starobinsky, Sov. Astron. Lett. 11, 133 (1985).
\bibitem{weinberg} S. Weinberg, Cosmology (Oxford University Press, New York, 2008).
 \bibitem{Dicus}D.Dicus and W.Repko, Phys.Rev.D ,72 ,088302(2005)[arXiv:0509096[astro-ph]]
 
\end{thebibliography}
\end{document}